\documentclass[fleqn,usenatbib]{mnras}

\usepackage{newtxtext,newtxmath}
\usepackage[T1]{fontenc}
\usepackage{ae,aecompl}

\usepackage{graphicx}	% Including figure files
\usepackage{amsmath}	% Advanced maths commands
\usepackage{amssymb}	% Extra maths symbols
\title[Spatially-Resolved Star Formation Main Sequence]{Understanding the Scatter in the Spatially-resolved Star Formation Main Sequence of Local Massive Spiral Galaxies}

\author[Abdurro'uf \& Masayuki Akiyama]{
Abdurro'uf,$^{1}$\thanks{E-mail: abdurrouf@astr.tohoku.ac.jp}
Masayuki Akiyama,$^{1}$
\\
$^{1}$Astronomical Institute, Tohoku University,  Aramaki, Aoba, Sendai 980-8578, Japan\\
}

\date{Accepted XXX. Received YYY; in original form ZZZ}

\pubyear{2017}

\begin{document}
\label{firstpage}
\pagerange{\pageref{firstpage}--\pageref{lastpage}}
\maketitle

\begin{abstract}
We investigate the relation between star formation rate (SFR) and stellar mass ($M_*$) in sub-galactic ($\sim 1$kpc) scale of 93 local ($0.01<z<0.02$) massive ($M_*>10^{10.5}M_{\odot}$) spiral galaxies. To derive spatially-resolved SFR and stellar mass, we perform so-called pixel-to-pixel SED fitting, which fits an observed spatially-resolved multiband SED with a library of model SEDs using Bayesian statistics approach. We use 2 bands (\textit{FUV} and \textit{NUV}) and 5 bands ($u$, $g$, $r$, $i$, and $z$) imaging data from Galaxy Evolution Explorer (GALEX) and Sloan Digital Sky Survey (SDSS), respectively. We find a tight nearly linear relation between the local surface density of SFR ($\Sigma_{\rm{SFR}}$) and stellar mass ($\Sigma_{*}$) which has flattening in high $\Sigma_{*}$. The near linear relation between $\Sigma_{*}$ and $\Sigma_{\rm SFR}$ suggests constant sSFR throughout the galaxies, and the scatter of the relation is directly related to that of sSFR. Therefore, we analyse the variation of sSFR in various scales. More massive galaxies on average have lower sSFR throughout them than less massive galaxies. We also find that barred galaxies have lower sSFR in a core region than non-barred galaxies. However, in the outside region, sSFR of barred and non-barred galaxies are similar and lead to the similar total sSFR.
\end{abstract}


%%%%%%%%%%%%%%%%%%%%%%%%%%%%%%%%%%%%%%%%%%%%%%%%%%

%%%%%%%%%%%%%%%%% BODY OF PAPER %%%%%%%%%%%%%%%%%%

\section{Introduction} \label{sec:intro}
It is already known for decades that among galaxies at narrow redshift range, there is a tight and nearly-linear relation between their $M_{*}$ and star formation rate (SFR), which is called star formation main sequence (SFMS)\citep[e.g.,][]{brinchmann2004, elbaz2007, noeske2007, salim2007, whitaker2012, speagle2014, renzini2015}. This relation means that current star formation rate and integrated past star formation rate are tracing each other tightly. Thanks to increasing number of wide-field and multiwavelength surveys, we can examine the relation with a large number of galaxies in a wide redshift range.

As galaxies evolve with redshift, global SFMS relation also evolves with redshift in term of its slope and normalization. The normalization shows a factor of $\sim 2$ dex decrease from high redshift ($z\sim 6$) to low redshift ($z\sim 0$), corresponds to decreasing specific SFR (sSFR) with cosmic time. On the other hand, its scatter seems to be independent of redshift with $\sigma$ of $\sim 0.3$ dex \citep{whitaker2012, speagle2014}. The constant small scatter of global SFMS relation may indicate that star formation inside galaxies is mainly governed by secular process rather than stochastic-merger process \citep{noeske2007}.

Recently, several investigation on local $\sim 1$kpc scale of star formation found that surface density of SFR ($\Sigma_{\rm SFR}$) and stellar mass ($\Sigma_{*}$) are correlated in the form similar as those in the global scale SFMS (\citealt{wuyts2013} (W13); \citealt{magdis2016} (M16); \citealt{canodiaz2016} (C16); \citealt{maragkoudakis2017} (M17)). These relations, which are called spatially-resolved SFMSs, are examined by comparing $\Sigma_{*}$ and $\Sigma_{\rm SFR}$ of sub-galactic regions of local $z\sim 0$ (C16 and M17) and $z\sim 1$ (W13 and M16) star-forming galaxies. Spatially-resolved SFMS provides the detailed information on the relation between the local star-formation mechanism and the global properties of galaxies.

Spatially-resolved SFMS relations reported in previous research papers differ in terms of slope and normalization. C16 found slope of $0.72$ and normalization of $-7.95$, while other three papers reported steeper slope close to unity with normalization of $\sim-8.3$ (W13 and M16) and $-9.01$ (M17). W13 and M16 seem to agree each other, while C16 and M17 result in different spatially-resolved SFMS. Similar to the global scale SFMS relations available in the literature, whose different slope, normalizations, and scatters (see \citet{speagle2014} for the compilation) are mainly caused by the use of different SFR indicators, initial mass function (IMF), stellar population synthesis model, and sample selection. The differences in the spatially-resolved SFMS relations are thought to be caused by differences in methodology (spatially-resolved spectroscopy in C16 and mid-infrared photometry in M17). Therefore, a unique method which can be applied to a large number of galaxies from low to high redshifts is needed to study spatially-resolved SFMS relation and its evolution with redshift.  

This work aims to investigate the local sub-galactic ($\sim 1$kpc) scale relation between $\Sigma_{*}$ and $\Sigma_{\rm SFR}$ using so-called pixel-to-pixel SED fitting, which fits spatially-resolved 7 bands (\textit{FUV}, \textit{NUV}, $u$, $g$, $r$, $i$, and $z$ from GALEX and SDSS) photometric SED with a library of model SEDs through Bayesian statistics approach. An advantage of this method is its applicability to various redshifts with nearly consistent manner. In subsequent papers, we will apply the same methodology, spatially-resolved rest-frame \textit{FUV}-\textit{NIR} SED fitting, to the galaxies at $z\sim 1$ to examine the evolution of the spatially-resolved SFMS relation.     

The structure of this paper is as follows. In Section~\ref{sec:data} we explain the data sample. Section~\ref{sec:metho} presents details of the methodology. Result and discussion are presented in Section~\ref{sec:result} and Section~\ref{sec:discussion}, respectively. Finally, we summarize our results in Section~\ref{sec:summary}. Throughout this paper, cosmological parameters of $\Omega_{m}=0.3$, $\Omega_{\Lambda}=0.7$, and $H_0=70 \text{kms}^{-1}\text{Mpc}^{-1}$ are used. $M_{*}$ will be used to represent total stellar masses of galaxies, while $m_{*}$ will represent the total stellar mass within a sub-galactic region. 

\section{Data Sample} \label{sec:data}

We examine the relation between $\Sigma_{*}$ and $\Sigma_{\rm SFR}$ in 1$-$2 kpc scale, 
by estimating spatially resolved $m_{*}$ and SFR of massive face-on spiral galaxies 
in the local universe. In order to estimate $\Sigma_{*}$ and $\Sigma_{\rm SFR}$, 
we use 2 (\textit{FUV} and \textit{NUV}) and 5 bands ($u$, $g$, $r$, $i$, and $z$) imaging data from GALEX \citep{morrissey2007} and SDSS DR10 \citep{ahn2014}, respectively. The 7 bands span from far-ultraviolet to near-infrared ($1350${\AA}$\le\lambda\le10000${\AA}). Thanks to the wide wavelength coverage, the degeneracy between age and dust extinction (existing in the stellar population synthesis modelings) can be broken; the dust extinction can be constrained by \textit{FUV}-\textit{NUV} colour, while age can be constrained by $u-g$ colour (which encloses the Balmer/4000{\AA} break).

In this analysis, we do not use MIR/FIR imaging dataset of nearby galaxies 
(e.g., The Spitzer Infrared Nearby Galaxies Survey (SINGS); \citet{kennicutt2003}, 
Key Insights on Nearby Galaxies: a Far-Infrared Survey with Herschel (KINGFISH); 
\citet{kennicutt2011}), 
because the numbers of massive galaxies in the above MIR/FIR imaging datasets 
are limited; for example, there are only 9 galaxies more massive 
than $M_{*}=10^{10.5} M_{\odot}$ in the KINGFISH sample \citep{kennicutt2011}.
Furthermore, in a subsequent paper, we will apply the same methodology, spatially-resolved 
rest-frame \textit{FUV}-\textit{NIR} SED fitting, to the galaxies at $z\sim1$ to examine the 
cosmological evolution of the $\Sigma_{*}$ and $\Sigma_{\rm SFR}$, and 
imaging dataset with 1 kpc scale resolution is not available in the MIR/FIR wavelength 
for the galaxies at $z\sim1$. 
 
The sample galaxies are selected from MPA-JHU (Max Planck Institute for 
Astrophysics-Johns Hopkins University) galaxy 
catalog\ \footnote{\url{http://www.mpa-garching.mpg.de/SDSS/DR7/}} \citep{tremonti2004} based on their redshifts and $M_{*}$. 
In the catalog, $M_{*}$ is calculated following a method used
in \citet{kauffmann2003} and \citet{salim2007}, and SFR is calculated 
based on emission line indices following \citet{brinchmann2004}. 
We apply the following criteria to the MPA-JHU galaxies that have 
GALEX imaging data; (1) redshift range of $0.01<z<0.02$, 
(2) $M_{*}$ more than $10^{10.5} M_{\odot}$, 
(3) spiral morphology, and (4) face-on disc with ellipticity 
less than 0.6 or $b/a > 0.4$.

\begin{figure}
\centering
\includegraphics[width=0.5\textwidth]{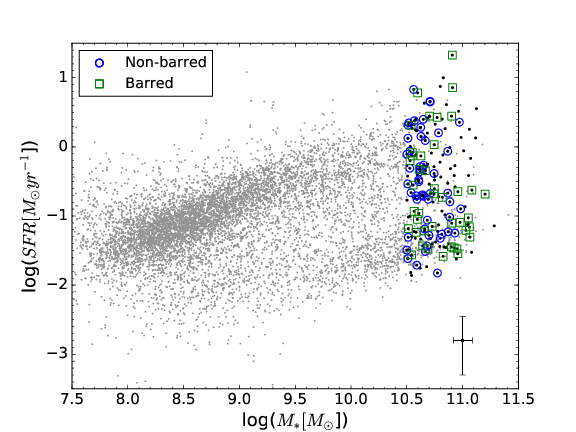}
\caption{SFR vs. $M_{*}$ of galaxies at $0.01<z<0.02$.
Grey dots represent all galaxies in the MPA-JHU galaxy 
catalog. Large black points represent spiral galaxies 
with $M_{*}>10^{10.5} M_{\odot}$, while the selected 
galaxies for this study are marked with blue open circles and
green open squares for non-barred and barred galaxies, respectively. Cross symbol in lower left side indicates average uncertainties of $M_{*}$ and SFR.
\label{fig:galaxies_sample}}
\end{figure}

The redshift range of the sample is determined,
in order to achieve the spatial resolution less than 1kpc with the GALEX
image, which has a FWHM of 5.3$^{\prime\prime}$ in the \textit{NUV} band. Because we focus
on the massive galaxies, the $M_{*}$ limit is set. 
We limit our analysis to spiral galaxies with face-on configuration 
in order to minimize the effect of dust extinction in highly inclined galaxies.
We check the morphological classification of the galaxies selected from 
the criteria (1) and (2) in Galaxyzoo\ \footnote{\url{www.galaxyzoo.org}} 
project \citep{lintott2008}, and also inspect the SDSS three colour
composite image with $g$, $r$, and $i$ bands by eye. 
Ellipticities of the selected spiral galaxies are calculated by averaging 
$r$-band elliptical-isophotes outside of the effective radius as
described in the construction of radial profiles 
(see Section \ref{subsec:radialprofiles}). In addition to the above
4 selection criteria, we remove galaxies that are not covered in 
the GALEX imaging surveys, and only consider galaxies that have 
more than 4 bins of pixels with S/N ratio more than 10 in all of the
7 bands (see Section~\ref{subsec:binning} for the binning method).
In total 93 galaxies are selected with the above criteria.
Figure~\ref{fig:galaxies_sample} shows the distribution of $M_{*}$ 
and SFR of the selected galaxies, which are marked with 
blue open circles and green open squares for 
non-barred and barred galaxies, respectively,
along with all galaxies at $0.01<z<0.02$ in the MPA-JHU galaxy 
catalog plotted with black and grey points.

In the later analysis, we consider the existence of bar 
structure in each spiral galaxy. For the classification of 
barred and non-barred galaxies, we follow the NASA/IPAC Extragalactic 
Database (NED)\footnote{\url{https://ned.ipac.caltech.edu/}}, where 
morphological classification of all selected galaxies are available.
Finally, the detailed morphological classification of the sample 
is as follows; the non-barred spiral sample contains 4 S0s, 13 spirals 
earlier type than Sb, and 34 spirals later type than Sb, and 
the barred spiral sample has 1 SB0, 15 spirals earlier than SBb, 
and 26 spirals later than SBb. 
All of the galaxies are earlier type than Scd, and there are no 
irregular galaxies.

The limiting magnitudes of the SDSS images for stellar objects are 
22.3, 23.3, 23.1, 22.3, and 20.8 mag in the $u$, $g$, $r$, $i$, and $z$ filters, 
respectively \citep{york2000}.
The depth and limiting magnitudes of GALEX images depend on 
the surveys; All Sky Survey (AIS) has exposure time of about 100 sec 
and limiting magnitudes ($5 \sigma$) of 19.9 mag in \textit{FUV} and 20.8 mag in \textit{NUV}, 
Medium Imaging Survey (MIS) has typical exposure time of 1500 sec and 
limiting magnitudes ($5 \sigma$) of 22.6 mag in \textit{FUV}, and 22.7 mag in \textit{NUV}, and 
Deep Imaging Survey (DIS) has typical exposure time of 30000 sec and 
limiting magnitudes ($5 \sigma$) of 24.8 mag in \textit{FUV} and 24.4 mag in \textit{NUV} \citep{morrissey2007}.
SDSS and GALEX images have a spatial sampling of 0.4$^{\prime\prime}$ per pixel 
and 1.5$^{\prime\prime}$ per pixel, respectively. 
The typical PSF sizes are 1.3$^{\prime\prime}$ for SDSS images, 
5.3$^{\prime\prime}$ for the GALEX \textit{NUV} band, and
4.2$^{\prime\prime}$ for the GALEX \textit{\textit{FUV}} band. The PSF sizes 
correspond to the physical sizes of 0.3$-$0.5kpc (SDSS images), 
1$-$2kpc (GALEX \textit{NUV} band), and 0.8$-$1.7 kpc (GALEX \textit{FUV}
band) for galaxies at $z=0.01\sim 0.02$. Individual pixel corresponds to the spatial scale of 0.3-0.6 kpc in the redshift range. 
%---------------------------------------------------         
%Figure :
%---------------------------------------------------

\section{Methodology} \label{sec:metho}

We apply a method so called "pixel-to-pixel SED fitting" \citep[e.g.,][]{abraham1999, welikala2008, zibetti2009, wuyts2012, sorba2015}
to derive $\Sigma_{*}$ and $\Sigma_{\rm SFR}$ in each region of a galaxy. 
This method can be divided into three main steps; (1) PSF matching,
image registration, and binning to obtain 7 bands photometric flux 
in each bin of a galaxy, 
(2) construction of model photometric SEDs, and (3) fitting the set of 
model photometric SEDs to the 7 band photometric flux. In this analysis, SED fitting based on Bayesian inference statistics is used to estimate $m_{*}$ and SFR of a bin. The Probability of each SED model, which has a corresponding $m_{*}$ and SFR, is calculated, then probability distribution functions (PDFs) of $m_{*}$ and SFR are constructed by compounding the probabilities of SED models in a library.

\subsection{PSF Matching and Image Registration}

The images in the \textit{FUV}, $u$, $g$, $r$, $i$, and $z$ bands are convolved
with the kernels to match the PSFs in these bands to that in the \textit{NUV} band,
which has the largest FWHM (5.3$^{\prime\prime}$) among the 7 bands.
The convolution kernels are calculated using \textbf{psfmatch} command 
in IRAF\ \footnote{IRAF is distributed by the National Optical Astronomy 
Observatory, which is operated by the Association of Universities for 
Research in Astronomy, Inc., under cooperative agreement with the National 
Science Foundation}, based on the average PSF of each band. 
The average PSF is determined by combining images of bright and 
isolated stars in the images. The average PSFs and kernels are calculated 
with the SDSS sampling ($0.4^{\prime\prime}$ per pixel), and the 
convolution is done under the sampling.

Then all the images are registered into the GALEX sampling ($1.5^{\prime\prime}$ per pixel). 
Afterward, the images are trimmed within 141$\times$141 pixels around the galaxy 
in question. The Fitting region of the galaxy is defined using segmentation maps 
produced by \textit{Sextractor} \citep{bertin1996} with a detection threshold of 
above 1.5 times larger than rms scatter outside of the galaxy applied to all of the 7 bands. The final fitting region of the galaxy is 
determined by joining all segmentation maps obtained in the 7 bands.

Flux and flux uncertainties in the 7 bands associated with each pixel 
are calculated by converting from pixel value at each band. All fluxes and flux 
uncertainties are expressed in $\text{erg} \ \text{s}^{-1} \text{cm}^{-2}${\AA}$^{-1}$. 
The conversion factors from pixel values of \textit{FUV} and \textit{NUV} GALEX images 
(in counts per second) to \textit{FUV}- and \textit{NUV}-band fluxes are explained in \citet{morrissey2007}. 
We convert pixel values of SDSS images, which are represented in unit of nanomaggy, 
to fluxes and flux uncertainties, following the description in the SDSS's 
website\ \footnote{\url{https://www.sdss3.org/dr10/algorithms/magnitudes.php} \\ 
\url{https://data.sdss.org/datamodel/files/BOSS_PHOTOOBJ/frames/RERUN/RUN/CAMCOL/frame.html}}.

\subsection{Pixel Binning} \label{subsec:binning}

In order to increase the S/N ratio of the resolved SED, we apply binning to the flux and flux uncertainties. We introduce a binning method in which three conditions
are considered; (1) closeness by which only connected pixels are binned, (2) similarity of SED's shape, and (3) binned S/N to achieve higher S/N in all bands than a threshold value. If we only consider the closeness of pixels' positions for the binning process, it is possible that some pixels may have different SED shapes compared to the other pixels in a bin. 

Firstly, the brightest pixel is selected from the trimmed $r$-band image of a galaxy. Then pixels enclosed within a radius of $r=2$ pixel from the brightest pixel are examined for similarity of it's SED shape to that of the brightest pixel. The similarity of the SED shape of a pixel with index of $m$ to that of the brightest pixel with index of $b$ is evaluated with $\chi^{2}$ of
%------------------------------------------------
% Equation (1): chi-square for binning
\begin{equation}
\chi^{2}=\sum_{i}\frac{(f_{m,i}-s_{mb}f_{b,i})^{2}}{\sigma_{m,i}^{2}+\sigma_{b,i}^{2}}.
\end{equation}
%-------------------------------------------------
$i$ in equation (1) represent image band, and $f_{m,i}$ and $f_{b,i}$ are $i$-th band flux of a pixel $m$ and $b$, respectively. $\sigma_{m,i}$ and $\sigma_{b,i}$ are $i$-th band flux uncertainty of the pixel $m$ and $b$. $s_{mb}$ corrects scaling difference between the two pixels, and it can be calculated using the equation of
%----------------------------------------- 
% Equation (2): normalization 
\begin{equation}
s_{mb}=\frac{\sum_{i}\frac{f_{m,i}f_{b,i}}{\sigma_{m,i}^{2}+\sigma_{b,i}^{2}}}{\sum_{i}\frac{f_{b,i}^{2}}{\sigma_{m,i}^{2}+\sigma_{b,i}^{2}}}.
\end{equation}
%-------------------------------------------
If $\chi^{2}$ is smaller than $30$, the pixel $m$ is binned with the brightest pixel $b$.

After binning all the selected pixels within the radius, the total S/N of the bin is checked. If total S/N of the bin is above the S/N threshold of S/N=10 in all of the 7 bands, the expansion of the bin is terminated. Otherwise, the radius of the process is increased by $\text{d}r=2$ pixel and the same procedure is applied until the total S/N of the bin reach the S/N threshold. Once the binning process of the bin is terminated, we look for the brightest pixel in the $r$-band image which has not been included in the previous binning and applies the same procedure again to make the next bin. 

The above procedure is applied until no bin can be made with remaining pixels. It is possible that some pixels of a galaxy, especially in the outskirt, are not included in any bin because shapes of their SEDs are different from those of neighbouring bins, and binning them with pixels with similar SEDs cannot reach the required S/N threshold. In such case, all of the remaining pixels are binned into one bin, finally.

\begin{figure}
\centering
\includegraphics[width=0.5\textwidth]{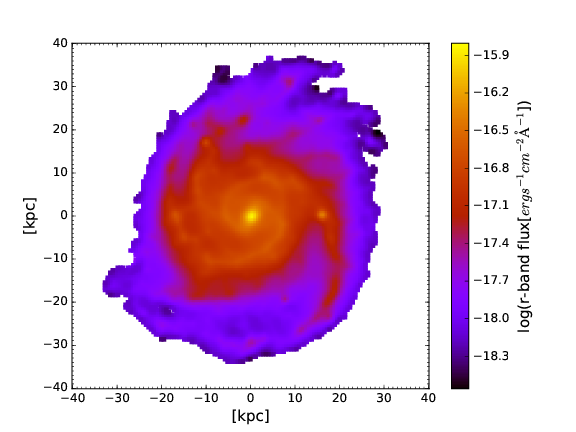}
\includegraphics[width=0.5\textwidth]{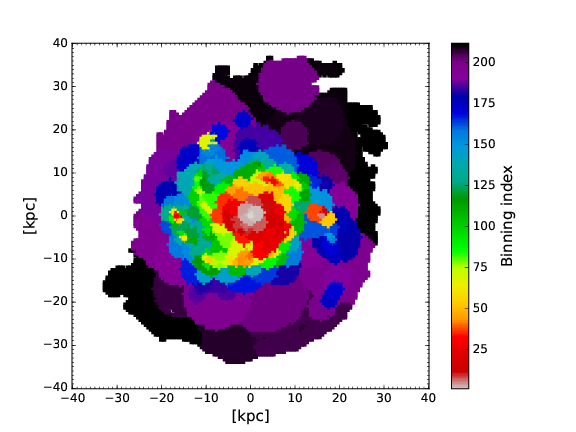}
\caption{Top panel: $r$-band image of SDSS ObjID 1237652947457998886 (or NGC0309). Bottom panel: the binning result of the galaxy. Different colour encoded with the index of a bin. 
North is to the top and east is to the left.
\label{fig:pixel_binning}}
\end{figure}

Figure~\ref{fig:pixel_binning} shows an example of the binning process of one of the sample galaxies (SDSS ObjID 1237652947457998886 or NGC309). The top panel shows its $r$-band image. As shown in the bottom panel, the galaxy has 212 bins shown with different colour encoded by the index of each bin.

%----------------------------------------------
% Figure : pixel binning result 
%----------------------------------------------

\subsection{Construction of a SED library}

The library of the SEDs used for the Bayesian analysis is constructed with the SED models with random parameters. In this analysis, we simply assume a uniform distribution within a certain parameter range, i.e. we do not consider any prior except for the minimum and maximum values.

\begin{figure*}
\centering
\includegraphics[width=0.98\textwidth]{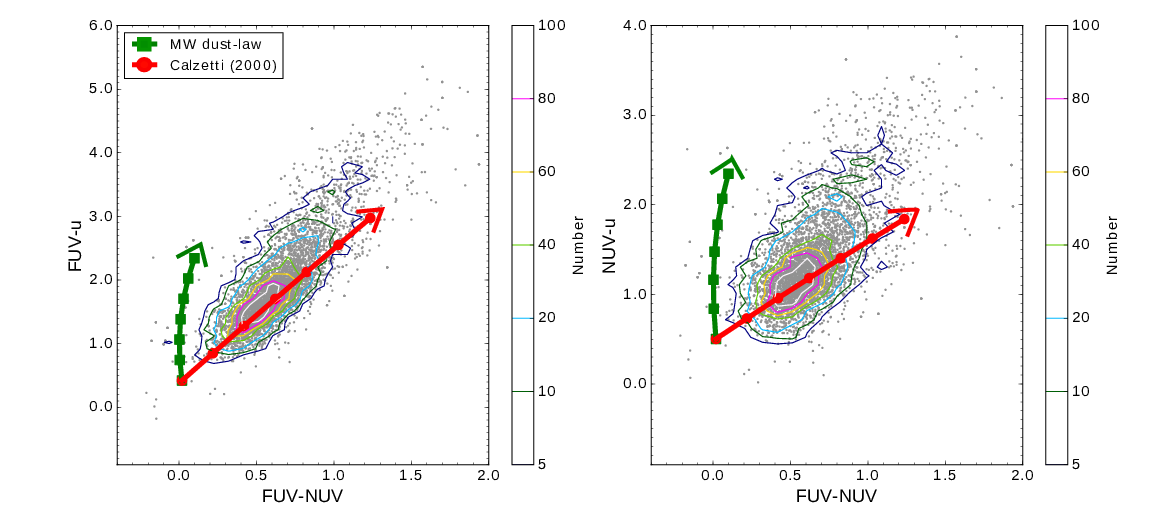}
\caption{Comparison between the dust extinction tracks of (left panel: \textit{FUV}$-$\textit{NUV} vs. \textit{FUV}$-u$, right panel: \textit{FUV}$-$\textit{NUV} vs. \textit{NUV}$-u$) Calzetti (2000) and Milky Way dust extinction laws for a model SED with $Z=0.02$, $\tau=8.6$Gyr, and age=$13.75$Gyr, in the colour-colour diagrams along with the distribution of the photometries of the bins in the galaxies. The circles and squares mark increasing colour excess $E(B-V)$ from 0 to 0.6 with step of 0.1.
\label{fig:Gals_dustcurve_test}}
\end{figure*}

Firstly, parent model spectra are generated using GALAXEV stellar population synthesis model \citep{bruzual2003} 
with uniform parameter grid. We assume \cite{chabrier2003} initial mass function (IMF) with mass range
between $0.1M_{\odot}$ and $100M_{\odot}$ and exponentially declining 
SFR model $\text{SFR}(t)\propto e^{-t/\tau}$ with $\tau$ of decaying time and $t$ of age. 
Dust extinction curve of \cite{calzetti2000} is applied as a screen because the extinction curve reproduces 
the observed SEDs better than other extinction curves. 
Figure~\ref{fig:Gals_dustcurve_test} compares the distribution of the bins of the spatially-resolved photometry of all of the galaxies in the sample (4768 bins) in the \textit{FUV}$-$\textit{NUV} vs. \textit{FUV}$-u$ and \textit{FUV}$-$\textit{NUV} vs. \textit{NUV}$-u$ colour-colour plane, and the vectors of the dust reddening effect with \citet{calzetti2000} and Milky-Way dust \citep{seaton1979, cardelli1989} extinction laws for a model SED with $Z=0.02$, $\tau=8.6$Gyr, and age=$13.75$Gyr. The model's parameters are selected as they are the typical parameters of best-fitting models with zero dust extinction. Milky-Way dust extinction law has the bump feature at around 2175{\AA}, thus reddening
effect in \textit{FUV}$-$\textit{NUV} colour is much smaller than that with the \citet{calzetti2000} extinction
law. Because the reddening vector with \citet{calzetti2000} extinction law broadly follows the 
observed distribution of the \textit{FUV}$-$\textit{NUV} vs. \textit{FUV}$-u$ and \textit{FUV}$-$\textit{NUV} vs. \textit{NUV}$-u$ colours of the bins better than 
the Milky-Way dust extinction law, we apply the \citet{calzetti2000} extinction law
in this study (see also Appendix~\ref{sec:robust}). 

We use the following grids of parameters in the parent model;
\begin{enumerate}
\item 100 values of SFR e-folding time, $\tau$, from 0.1 Gyr to 10 Gyr with $\bigtriangleup\tau$ of 0.1 Gyr,
\item 55 values of age, $t$, from 0.25 Gyr to 13.75 Gyr with $\bigtriangleup t$ of 0.25 Gyr,  
\item 13 values of $E(B-V)$, 0 (no extinction), 0.05, 0.1, 0.15, 0.2, 0.25, 0.3, 0.35, 0.4, 0.45, 0.5, 0.55, and 0.6 mag,
\item 4 values of metallicities, Z : 0.004, 0.008, 0.02, and 0.05.
\end{enumerate} 
Then, the model spectrum, which is normalized to the total stellar mass of 1 $M_{\odot}$, 
is converted to the observed flux, $f_{\lambda}(\lambda)$ with the 
cosmological redshifting and dimming effect. Fluxes of the models 
in the 7 bands are calculated using
%--------------------------------------------------- 
%equation 4 : integrating through filter
\begin{equation}
f_{m,i}=\frac{\int f_{\lambda}(\lambda)\lambda T_{i}(\lambda)d\lambda}{\int \lambda T_{i}(\lambda)d\lambda},
\end{equation}
%---------------------------------------------------
where $T_i$ is the $i$th-band filter transmission function and $f_{m,i}$ is the model's observed-frame flux density through that filter. 
The $\lambda$ factor in both of the numerator and denominator accounts for the fact that the observations are done with photon counting
detectors and the number of photons per unit flux are proportional to the wavelength \citep{sawicki2012}.

We construct 300,000 photometry models with a random set of parameters ($\tau$, $t$, $E(B-V)$, and $Z$) by interpolating the 286,000 parent model spectra. The sets of parameters are randomly selected in the linear scale within the same ranges as those of the parent models. For each set of parameters, 7 band fluxes and stellar mass are calculated by interpolating at first in 3 dimensional space of ($E(B-V)$, $t$, and $\tau$) using tricubic interpolation for each metalicity, then in 1-dimensional space of $Z$ with cubic spline interpolation. 

%------------------------------------
%------------------------------------
\subsection{\texorpdfstring{$m_{*}$}{Lg} and SFR from the SED fitting with Bayesian statistics}

$m_{*}$ and SFR of each bin of a galaxy are estimated with the Bayesian statistics.
At first, in each bin probability distribution functions (PDFs) of SFR and $m_{*}$ are constructed, 
and then the posterior means of SFR and $m_{*}$ are calculated. 
Because the model SEDs are normalized with a stellar mass of 1 $M_{\odot}$, 
the estimated normalization corresponds to the $m_{*}$. Instantaneous SFR is derived with 
%---------------------------------------
% Equation : SFR 
\begin{equation}
SFR=\frac{m_{*}}{1-\exp(-t/\tau)} \times \frac{\exp(-t/\tau)}{\tau}
\end{equation}
%---------------------------------------
where first multiplication term in the right-hand side serve as scaling between 
integrated mass and instantaneous SFR. 
The probability functions are constructed based on the $\chi^{2}$ statistics. 
The $\chi^{2}$ value of the $m$-th model for the SED of a $d$-th bin is defined with  
%--------------------------------------------------------------
% Equation 4: chi-square
\begin{equation}
\chi^{2}_{m}=\sum_{i}(\frac{f_{d,i}-sf_{m,i}}{\sigma_{i}})^{2},
\end{equation} 
%------------------------------------------------------
where $s$ represents the scaling of the model SED. The scaling factor $s$ is randomly drawn between $0.1 \times s_{\rm least}$ to $10 \times s_{\rm least}$, where $s_{\rm least}$ is the scaling of the minimum $\chi^{2}$ solution, that is
%----------------------------------------------------------
% Equation 5 : least-square s 
\begin{equation}
s_{\rm least}=\frac{\sum\limits_{i}\frac{f_{d,i}f_{m,i}}{\sigma_{i}^{2}}}{\sum\limits_{i}\frac{f_{m,i}^{2}}{\sigma_{i}^{2}}}.
\end{equation}        
%------------------------------------------------------
For each model SED, 100 random $s$ are generated and $\chi^{2}$ of model SED with each $s$ is calculated.

The posterior mean of the SFR and $m_*$ of a bin are calculated by 
%---------------------------------------------------
% Equation 6 : posterior mean 
\begin{equation}
\overline{\theta}=\int \theta P(\theta|X)d\theta,
\end{equation}
%---------------------------------------------------
with $\theta$ stands for SFR or $m_*$, 
and $P(\theta|X)$ is the probability of a model with a 
value of $\theta$ given an observational data set, $X$.

We use a PDF in the form of the Student's t-distribution, which gives 
higher probability to a model with large $\chi^{2}$ compared to the 
Gaussian distribution ($P(\theta|X)\propto \exp(-\chi^{2}/2)$), 
which is commonly used in the Bayesian SED 
fittings \citep[e.g.,][]{kauffmann2003, salim2007, dacunha2008, walcher2011, hanhan2014}.
We assume the degree of freedom of the Student's t-distribution, 
$\nu$, as 3, because that distribution gives the consistent sSFR estimate 
as tested with mock SEDs and also gives the best SFR estimate as compared to 
24$\mu$m observations (see Appendix~\ref{sec:robust}).
The normalized PDF has a form as
%---------------------------------------------------
% Equation 7 : probability
\begin{equation}
P(\theta|X)=\frac{W_{\theta m}}{\sum\limits_{m=1}^{N}W_{\theta m}}
\end{equation}
with 
% Equation 8 : weight
\begin{equation}
W_{\theta m}=\frac{\Gamma\left(\frac{\nu+1}{2}\right)}{\sqrt{\nu\pi}\Gamma\left(\frac{\nu}{2}\right)}\left(1+\frac{(\chi_{m}^{2})^{2}}{\nu}\right)^{-\frac{\nu+1}{2}}
\end{equation} 
%---------------------------------------------------
with $W_{\theta m}$ represents the weight of the $m$-th SED model which 
has $\theta$ (SFR or $m_{*}$), and $N$ is the total number of 
the SED models, which is 100$\times$300,000.

Once the posterior means of the SFR and $m_{*}$ of each bin 
are calculated, those values are divided into the pixels which 
belong to the bin. We assume that the SFR and $m_{*}$ of a pixel 
are proportional to its \textit{FUV} and $z$ band flux, respectively. We also calculate uncertainties of SFR and $m_*$ estimates by 
calculating the standard deviation with the PDFs of 
SFR and $m_{*}$.

\section{Results} \label{sec:result}

\subsection{Spatially-resolved Star Formation Main Sequence}
\label{subsec:spatially_resolved_SFMS}

\begin{figure*}
\centering
\includegraphics[width=1\textwidth]{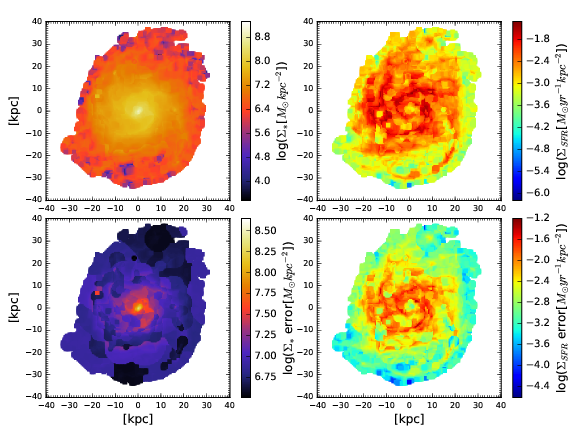}
\caption{An example of the pixel-to-pixel SED fitting of a galaxy (SDSS ObjID 1237652947457998886 or NGC0309). Top-left panel: $m_{*}$ surface density map, top-right panel: SFR surface density map, bottom-left panel: uncertainty of $m_{*}$ surface density map, and bottom-right panel: uncertainty of the SFR surface density map. Up and left directions correspond to the north and east respectively.
\label{fig:p2p_example}}
\end{figure*}

Figure~\ref{fig:p2p_example} shows pixel-to-pixel SED fitting 
result of a galaxy in the sample (SDSS ObjID 1237652947457998886 or 
NGC0309, whose pixel binning result is shown 
in Figure~\ref{fig:pixel_binning}). The SFR surface density, 
$\Sigma_{\rm SFR}$, map clearly traces spiral arm pattern and 
some knots which enhanced star formation activities are 
associated, while the $m_{*}$ surface density, $\Sigma_{*}$, map 
shows smoother distribution.

%-----------------------------------------------------
% Figure : Example of fitting result
%-----------------------------------------------------

\begin{figure}
\centering
\includegraphics[width=0.5\textwidth]{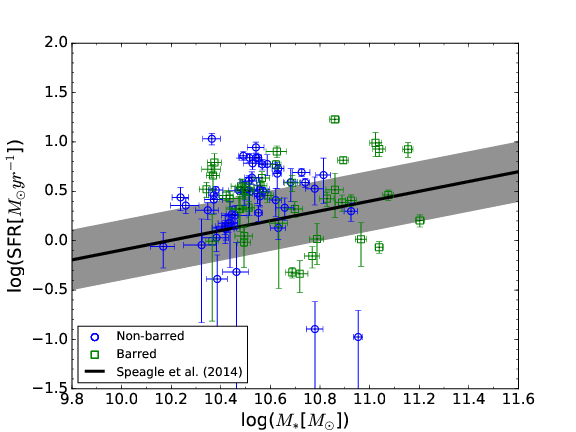}
\caption{Relation between the integrated $M_{*}$ and SFRs. 
Blue open circles and green open squares represent 
non-barred and barred galaxies, respectively.
The Black line represents integrated SFMS relation of \citet{speagle2014}. 
Grey-shaded region around the line represents $\pm 0.3$ dex scatter.
\label{fig:integrated_SFMS}}
\end{figure}

Figure~\ref{fig:integrated_SFMS} shows integrated $M_{*}$ vs. SFR of the 93 galaxies 
in the sample. The black line represents integrated star formation main sequence (SFMS)
of \citet{speagle2014} (calculated at the median redshift of the sample, $z=0.0165$) 
with $\pm 0.3$ dex scatter shown by grey-shaded region.
This integrated SFMS is obtained by calculating the total stellar mass, $M_{*}$, 
and SFR of each galaxy by summing up those values within the fitting region. The uncertainties associated with $M_{*}$ and SFR are calculated by propagation error of summation. The points are distributed around the black line.
%---------------------------------------
% Figure : integrated SFMS
%---------------------------------------

\begin{figure*}
\centering
\includegraphics[width=0.99\textwidth]{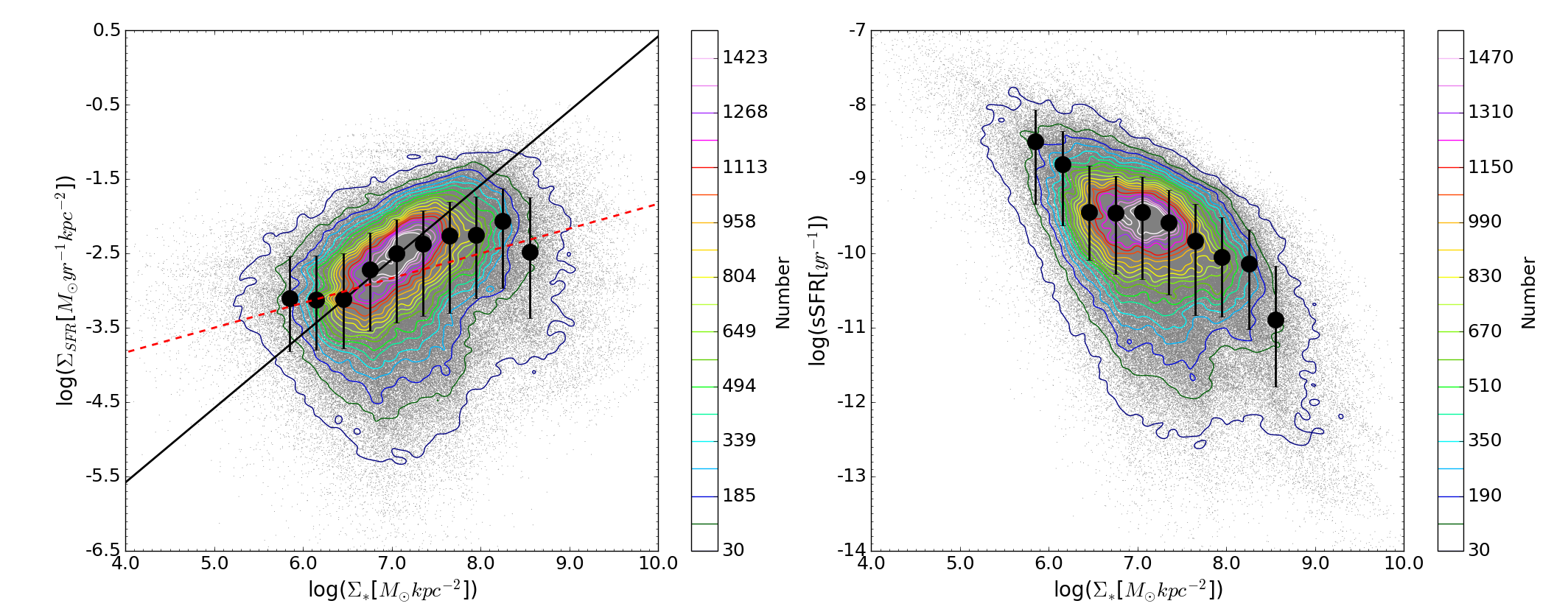}
\caption{Left panel: Distribution of surface $M_{*}$ and SFR density of the pixels along with contours colour coded by number in each bin. Red dashed line represents linear function fitting to the data. Black circles with errorbars show the mode of $\Sigma_{\rm SFR}$ distribution for $\Sigma_{*}$ bin with a 0.3 dex width. The Black solid line represents a linear function fitting to the linear increasing part of the mode values. Right panel: Distribution of $\Sigma_{*}$ vs. sSFR. Black circles with errorbars show modes of sSFR distribution for $\Sigma_{*}$ bin with 0.3 dex width. 
\label{fig:resolved_SFMS_pix}}
\end{figure*}

To examine the relation between $\Sigma_{*}$ and $\Sigma_{\rm SFR}$ in $\le$ 1-2 kpc scale, the $\Sigma_{*}$ and $\Sigma_{\rm SFR}$ of all 375,215 pixels of the sample galaxies are plotted in the left panel of Figure~\ref{fig:resolved_SFMS_pix}. The figure shows scatter of the pixels and contour colour-coded with the number in each $0.1\times0.1$ dex bin. As shown by the contour with a high number density, there is a tight relation between $\Sigma_{*}$ and $\Sigma_{\rm SFR}$. Fitting the data with a linear relation with the form of 
%----------------------------------------------------
% Equation : linear fitting SFMS
\begin{equation}
\log \Sigma_{\rm SFR}=\alpha\log \Sigma_{*}+\beta.
\end{equation}          
%-----------------------------------------------------
through least-square fitting method, resulted in the best-fitting relation, which is shown with red dashed line and has a slope ($\alpha$) of $0.33$ and zero point ($\beta$) of $-5.23$ with a scatter ($\sigma$) of $0.70$ dex. The best-fitting relation looks largely deviate from overall data distribution shown by contours which is caused by large scatter of high-mass pixels with low sSFR. Therefore, a more objective method is needed to calculate the slope and zero point of the main-sequence. \citet{renzini2015} proposed an objective way to define the main-sequence by calculating mode at "fixed $M_{*}$" which corresponds to the peak of SFR distribution at fixed $M_{*}$. Although the method is developed for the integrated SFMS, but it is still relevant for the spatially-resolved SFMS. Black circles in the left panel of Figure~\ref{fig:resolved_SFMS_pix} show mode of $\Sigma_{\rm SFR}$ distribution for $\Sigma_{*}$ bin with 0.3 dex width. Errorbars represent standard deviation from the mode and calculated separately above and below the mode value. The mode values as a function of $\Sigma_{*}$ seem to well follow the distribution of the data shown by the contours. Steepening (flattening) in lower (higher) than $\log\Sigma_{*}\sim7.5$ is clearly shown by the mode values, while one should note that flattening in two lowest mode values is affected by the detection limit. 

In the right panel of Figure~\ref{fig:resolved_SFMS_pix}, sSFR, which is a ratio between $\Sigma_{\rm SFR}$ and $\Sigma_{*}$, is plotted as a function of $\Sigma_{*}$. The black circles show the mode of sSFR distribution for a $\Sigma_{*}$ bin with 0.3 dex width. The errorbar is calculated in the same way as those in the left panel. The general trend shows decline of sSFR with $\Sigma_{*}$ with a flat profile in the $\Sigma_{*}$ range, $10^{6.15}M_\odot\text{kpc}^{-2}\lesssim \Sigma_{*} \lesssim 10^{7.65}M_\odot\text{kpc}^{-2}$, which correspond to linear increase in the same $\Sigma_{*}$ range in the left panel. Applying linear function fitting to the four mode values enclosed within the red contour, resulted in steeper slope ($\alpha$), $1.00$ with zero point ($\beta$) of $-9.58$, which represented by the black solid line. Firstly, the zero point ($\beta$) is determined by the average $\log(\rm sSFR)$ of the four values, then the best slope ($\alpha$) is determined by a fitting with equation (10). The black solid line well represents the distribution of the data shown by the contours especially in the peak of the distribution. 

It is possible that this result has implication on the nearly proportionality relation between an integrated SFR and a total stellar mass of a disc, $M_{*,\rm{disc}}$, without a bulge component, of the local star-forming galaxies observed by \citet{abramson2014}.

\begin{figure*}
\centering
\includegraphics[width=0.99\textwidth]{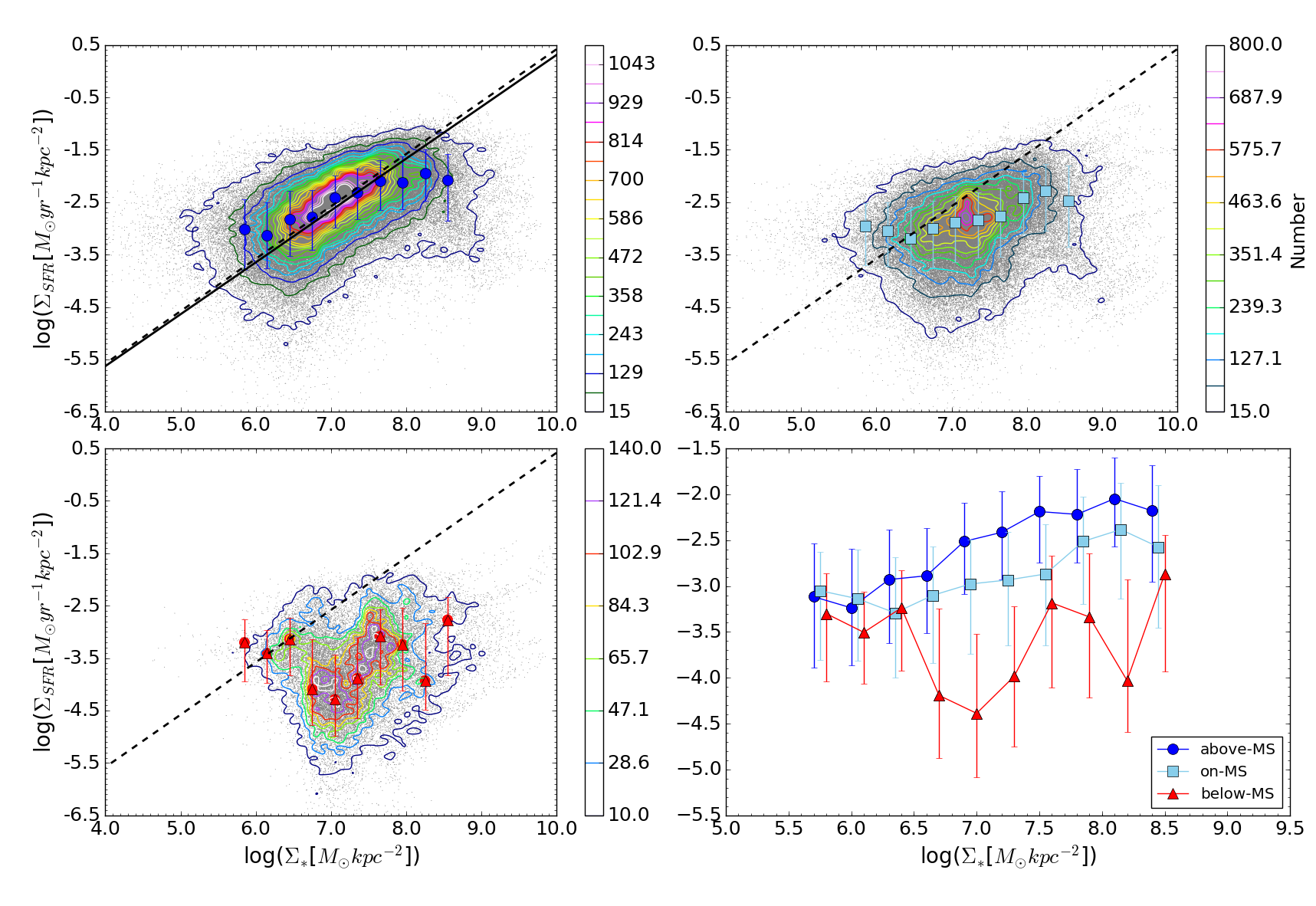}
\caption{Spatially-resolved SFMS from galaxies above the main-sequence (top-left panel), on the main-sequence (top-right panel), and below the main-sequence (bottom-left panel). Global main-sequence used here is following \citet{speagle2014} (see Figure~\ref{fig:integrated_SFMS}). Black solid line in the top-left panel represents a linear function fitting to the linear increasing part of the mode values. Black dashed lines in three panels are the same as the black solid line in the left panel of Figure~\ref{fig:resolved_SFMS_pix}. Bottom-right panel: Comparison between the mode profiles of all the three groups. For clarity, blue circles are shifted by 0.05 dex to the left, while red triangles are shifted by 0.05 dex to the right from their actual positions.}
\label{fig:resolved_SFMS_SBSFQS}
\end{figure*}

Recently reported spatially-resolved SFMS in the literature (W13, M16, C16, and M17) are all defined with star-forming galaxy sample. Here, our sample contains some galaxies lie far below the main-sequence (see Figure~\ref{fig:integrated_SFMS}) which seem to be quenched galaxies. These galaxies have lower spatially-resolved sSFR than those galaxies in the main-sequence (MS). This is one of the contributors to the large scatter of the spatially-resolved SFMS derived in this work. Therefore, here we separate galaxies sample into three groups based on their position relative to the SFMS relation of \citet{speagle2014}: above-MS (above $+0.3$ dex, contains 50 galaxies), on-MS (within $\pm 0.3$ dex, contains 34 galaxies), and below-MS (below $-0.3$ dex, contains 9 galaxies). Figure~\ref{fig:resolved_SFMS_SBSFQS} shows spatially-resolved SFMS of galaxies above-MS (top-left, contains 218593 pixels), on-MS (top-right, contains 120161 pixels), and below-MS (bottom-left, contains 36461 pixels). Spatially-resolved SFMS is clearly shown in the first group (above-MS) with steepening (flattening) in low (high) $\Sigma_{*}$ similar as those in the left panel of Figure~\ref{fig:resolved_SFMS_pix} (for all galaxies). The best-fitting linear function (represented by the black solid line) to the linear increasing part of the mode values (consist of four mode values), calculated with the same method as for Figure~\ref{fig:resolved_SFMS_pix}, has a slope of $0.99$ and zero point of $-9.59$. Furthermore, linear function fitting to all scattered data in this group resulted in lower scatter, ($\sigma$) of $0.55$, than those for all galaxies. The black dashed line in each panel is the same as the black solid line in the left panel of Figure~\ref{fig:resolved_SFMS_pix}. In the bottom right panel, mode profiles of the three groups are compared, which shows systematically different normalization (which corresponds to the difference in sSFR) between the three groups. There is no clear trend of spatially-resolved SFMS in the quenched galaxies as shown in the bottom-left panel.

%----------------------------------------------
% Figure : resolved SFMS
%-------------------------------
%------------------------------------------------
\begin{figure}
\includegraphics[width=0.5\textwidth]{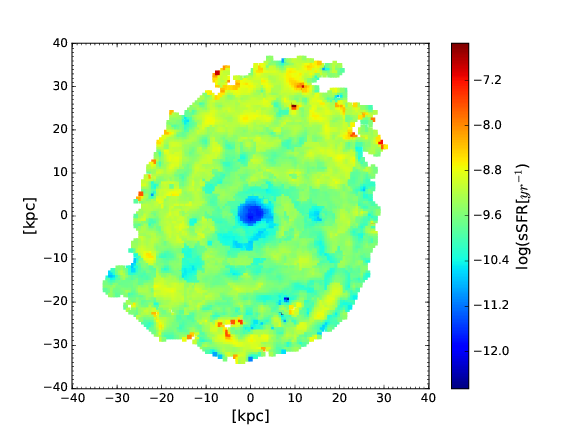}
\caption{Spatial variation of the sSFR in galaxy SDSS ObjID 1237652947457998886 or 
NGC0309.
\label{fig:Gal9_sSFRmap}}
\end{figure}

%------------------------------------------------  
The systematic deviation from a constant sSFR implies systematically different spatially-resolved SFMS in the location. Therefore, hereafter, we examine the systematic offset of sSFR in various scale. At first, Figure~\ref{fig:Gal9_sSFRmap} shows sSFR map of the galaxy SDSS ObjID 1237652947457998886 or NGC0309. The spiral arms structure is clearly shown on the map indicating the enhanced sSFR along the spiral arms compare to the underlying disc region. Therefore, on the smallest scale, the spatially-resolved SFMS in the spiral arm regions has higher normalization than that in the underlying disc region.      
%-----------------------------------------
% Figure :
%-------------------------------------

\subsection{Radial Profiles of \texorpdfstring{$\Sigma_{\rm SFR}$}{Lg}, \texorpdfstring{$\Sigma_{*}$}{Lg}, and sSFR} \label{subsec:radialprofiles}

\begin{figure}
\centering
\includegraphics[width=0.5\textwidth]{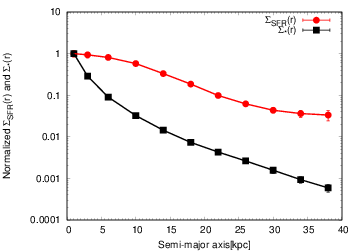}
\caption{Comparison between normalized average $\Sigma_{*}$ (black filled square) and $\Sigma_{\rm SFR}$ (red filled circle) radial profiles. The radial profiles are constructed by calculating average of the radial profiles then normalized the average radial profiles by their central values. The errorbars are calculated from standard error of mean.}
\label{fig:average_radprof_SFRSM}
\end{figure}

In order to examine the radial variation of the spatially-resolved SFMS, we construct the radial profiles of the sSFR. 
Firstly, radial profiles of $\Sigma_{*}$ and $\Sigma_{\rm SFR}$ of each galaxy are constructed as a function of the semi-major axis of ellipsoids. The ellipsoids are determined by applying elliptical-isophotes fitting to the $r$-band images of the galaxies using \textbf{ellipse} command in IRAF. Then average ellipticity and position angle are derived using ellipsoids outside of the half-mass radius of a galaxy. The half-mass radius is defined with the length of semi-major axis 
enclosing half of the total $M_{*}$ calculated with average ellipticity and position angle in the entire radius range. We only consider ellipsoids in the outer region of a galaxy, in order to escape from an effect of an existence of bulge and bar components. The radial profile is sampled in 2 kpc step 
within 6 kpc, and 4 kpc step outside of it. We extrapolate the radial profiles of each galaxy up to 40 kpc by fitting the exponential law outside of 6 kpc as follows,
%-----------------------------
% Equation : exponential function :
\begin{equation}
\psi(r)=\psi_{0}e^{r/h}
\end{equation}
%-----------------------------
with $\psi$ stand for $M_*$ or SFR. $\psi_{0}$ and $h$ represent zero point and radial scale-length, respectively. Figure~\ref{fig:average_radprof_SFRSM} shows normalized average radial profiles of $\Sigma_{*}$ and $\Sigma_{\rm SFR}$ which are constructed by calculating the average of the radial profiles, then normalize the average radial profiles with their central values. The normalized average radial profile of $\Sigma_{\rm SFR}$ has shallower and more extended profile than the average radial profile of $\Sigma_{*}$ which shows more centrally concentrated distribution.

%-----------------------------------------------------------
% Figure : SFR vs SM radial profile
%---------------------------------------------------------

\begin{figure}
\includegraphics[width=0.5\textwidth]{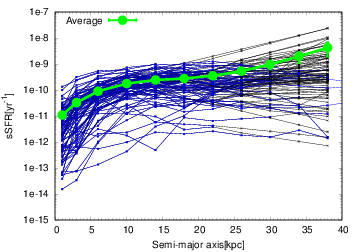}
\caption{sSFR($r$) profiles of 93 galaxies in the data sample (shown with blue filled square) with extrapolated part of the profile (shown with black cross) and their average (shown with green colour). The error bars are calculated from standard error of mean.  
\label{fig:plot_average_sSFR_radprof}}
\end{figure}

Utilizing the individual and averaged radial profiles of $\Sigma_{*}$ and $\Sigma_{\rm SFR}$, we derive profiles of sSFR. Figure~\ref{fig:plot_average_sSFR_radprof} shows
sSFR($r$) profiles of all of the 93 galaxies. Black cross 
represent extrapolated part of the profile. The average profile is shown with green filled circles. The average radial profile of sSFR is increasing as increasing radius with stronger increase found around the central region ($r \gtrsim 6$ kpc). The profile is nearly flat within the radius range of $10 \text{kpc} \leq r \leq 22 \text{kpc}$, which agree with the existence of spatially-resolved SFMS (see Section~\ref{subsec:spatially_resolved_SFMS}). The sSFR in the central region has on average 0.9 dex smaller than those in the outer disc region. 
Such decline is consistent with an existence of a bulge component with an old stellar population and with the recent results with integral field spectroscopy \citep{gonzalez2016}. The Inverse of the sSFR corresponds to the time scale of the star-formation. The average radial profile of the sSFR of the sample show lower than $10^{-10}$ yr$^{-1}$ in the central $r \lesssim 6$ kpc region but higher than $10^{-10}$ yr$^{-1}$ outside of it ($r > 6$ kpc). This critical sSFR corresponds to the Hubble time.

%-----------------------------------------------------------
% Figure : sSFR radial profile
%-----------------------------------------------------------

\section{Discussion} \label{sec:discussion}

\subsection{Radial variation of the spatially-resolved SFMS}

\begin{figure*}
\centering
\includegraphics[width=0.7\textwidth]{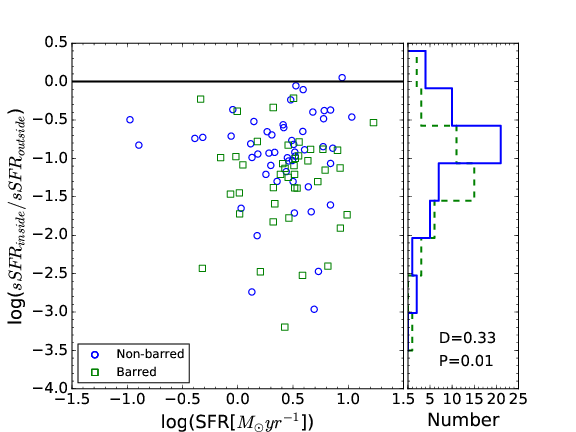}
\caption{Total SFR vs. logarithmic ratio of $\rm SFR_{\rm inside}/\rm SFR_{\rm outside}$ of the sample galaxies. Barred and non-barred galaxies are shown with green open squares and blue open circles, respectively. Right panel shows the distribuions of the ratio for barred (green dashed-line) and non-barred (blue solid-line) galaxies.
\label{fig:TotSFR_vs_ratSFRinSFRout}}
\end{figure*}

In this discussion, 
we focus on the variation of spatially-resolved SFMS in large scale, 
and we evaluate the systematic variation utilizing the radial dependence of
the sSFR. In order to quantify the difference in the profiles of sSFR among 
the galaxies, we derive the ratio between the integrated sSFR inside and outside 
of the half-mass radius ($r_{e,M_{*}}$) same as those defined in Section~\ref{subsec:radialprofiles}. If the ratio becomes larger (smaller) than 1, we expect sSFR is higher in the inner (outer) region. Figure~\ref{fig:TotSFR_vs_ratSFRinSFRout} shows the ratios as a 
function of the total SFR. As expected from the radial profiles of $\Sigma_{*}$ 
and $\Sigma_{\rm SFR}$, most of the galaxies have the ratio smaller than 1. 

By checking the morphologies of the galaxies as a function of the ratio, we found that galaxies that have small ratios are dominated by barred galaxies. In the Figure~\ref{fig:TotSFR_vs_ratSFRinSFRout}, barred and non-barred galaxies are shown with green squares and blue circles, respectively. The barred galaxies have smaller ratios on average, median values of barred and non-barred galaxies are $-1.08$ and $-0.78$ dex, respectively. The KS-test of the difference in the distributions of the two samples implies that they are different with a significance level of 0.05. Statistical D- and P-values of the KS-test are shown in the histogram of Figure~\ref{fig:TotSFR_vs_ratSFRinSFRout}. The results indicate that the star-formation of the barred galaxies is more suppressed in the inner region compared to the non-barred galaxies. 

%-----------------------------------------------------------
% Figure : SFR_inside vs SFR_outside
%-----------------------------------------------------------  

\begin{figure*}
\centering
\includegraphics[width=1\textwidth]{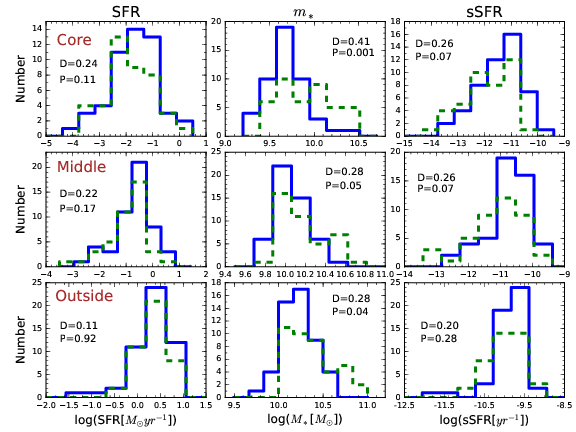}
\caption{Distributions of SFR, $m_*$, and sSFR of barred- and non-barred galaxies in three regions: "core" defined as region inside $10^{-0.5}r_e(M_*)$, "middle" defined as $10^{-0.5}r_e(M_*)<r\leqslant r_e(M_*)$, and "outside" defined as $r>r_e(M_*)$. 
The histograms are shown along with D- and P- values resulted from KS test. Solid blue and dashed green histograms are for non-barred and barred galaxies, respectively.
\label{fig:SFMS_coremiddleout}}
\end{figure*}

In order to further examine the difference in the barred and non-barred galaxies, we divide the galaxies into three regions: "core" defined as region inside $10^{-0.5} r_{e,M_{*}}$, "middle" defined as $10^{-0.5}r_{e,M_{*}}<r\leqslant r_{e,M_{*}}$, and "outside" defined as $r>r_{e,M_{*}}$ following \citet{wuyts2013}. 
Figure~\ref{fig:SFMS_coremiddleout} shows the distributions of summed SFRs, $m_*$, and sSFRs in the core, middle, and outside regions. 
As can be seen in the histograms, barred- and non-barred galaxies
have similar SFRs each other in the three regions. The $m_{*}$ distribution of barred- and non-barred galaxies are significantly different in the core region, where barred galaxies have higher $m_{*}$ than non-barred galaxies, while the difference in the middle and outside regions are small. sSFR distribution of barred- and non-barred galaxies are slightly different in core and middle, where barred-galaxies have lower median than non-barred galaxies, while in the outer region they have similar value.

%-------------------------------------------
%Figure : local variation of SFR, SM ,and sSFR     
%-------------------------------------------------

Global morphology of a galaxy is affecting the
spatially-resolved SFMS, but limited to the 
core of a galaxy. In the core region, barred galaxies
have lower sSFR than non-barred galaxies. 
It has been suggested that a bar structure can 
quench star formation of galaxies
\citep{cheung2013, gavazzi2015, zhou2015, spinoso2017}.
Numerical simulation of a bar formation in a
galaxy shows that gas is rapidly transferred toward the
central region on the onset of the bar structure, and
that the gas transfer triggers a burst of star formation
in the central region of the galaxy. As the gas being 
transferred to the central region, gas within the
bar's corotation radius is consumed and SFR
within the radius can be suppressed finally \citep{spinoso2017}.

Interestingly, such suppression of the sSFR 
is only seen in the core region and there is no
difference of sSFR in the outer region between
barred and non-barred galaxies. The total 
sSFRs of barred and non-barred galaxies do
not show a significant difference, and this opposes
the arguments that bar structure can quench the
star formation activity in the entire region.
The similarity of the distributions of sSFR in the
middle and outside regions suggests that 
the spatially-resolved SFMS does not vary in the 
outside of the core region, and does not depend on 
the existence of a bar structure in the central region of a disc.

\subsection{The Dependence of sSFR(\texorpdfstring{$r$}{Lg}) Profile on Integrated \texorpdfstring{$M_*$}{Lg}}

\begin{figure}
\centering
\includegraphics[width=0.5\textwidth]{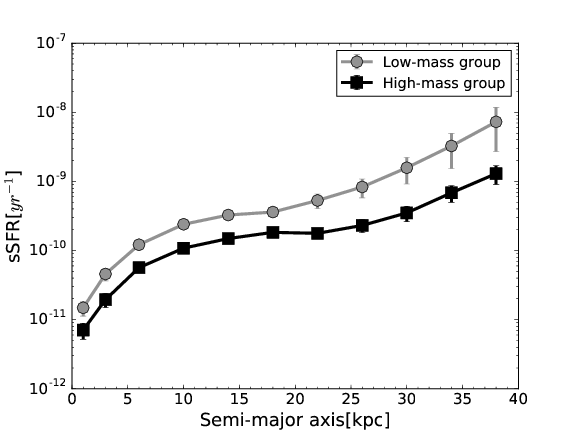}
\caption{Comparison between the average sSFR($r$) profiles of low- and high-mass groups (grey circle and black square, respectively) in the galaxy sample. The error bars are derived from standard error of mean.
\label{fig:average_radprof_sSFR_highlowmass}}
\end{figure}

In order to examine the dependence of spatially-resolved SFMS on 
the global property of galaxies, we divide the sample into two
groups based on their $M_{*}$ at the median $M_{*}$ of $3.5\times 10^{10}M_{\odot}$.
Figure~\ref{fig:average_radprof_sSFR_highlowmass} shows the comparison
between sSFR($r$) profiles of the high- and low-mass groups.
Both of the profiles are increasing as increasing radius and show sharp decline in the central region. These shapes are similar to the average of the entire sample shown 
in Figure~\ref{fig:plot_average_sSFR_radprof}.
In all radius, the sSFR($r$) of the low-mass group is higher
than the high-mass group. 

\begin{figure}
\centering
\includegraphics[width=0.5\textwidth]{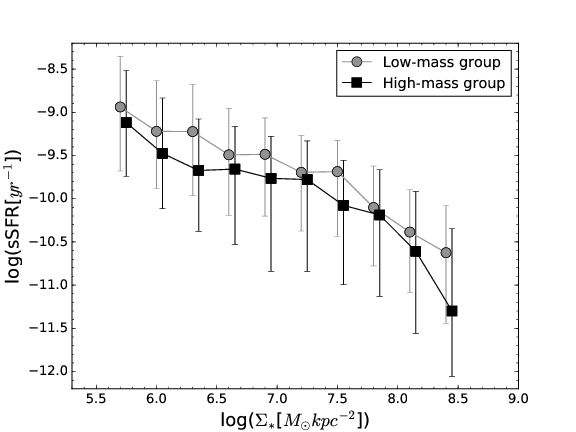}
\caption{Comparison between the mode sSFR vs. $\Sigma_*$ profiles of low- and high-mass groups (grey circle and black square, respectively) in the galaxy sample. For clarity, grey circles are shifted by 0.05 dex to the left.
\label{fig:sSFRvsSM_lessmoremassgals}}
\end{figure}

Figure~\ref{fig:sSFRvsSM_lessmoremassgals} shows comparison between 
mode values of sSFR in each 0.3 dex bin of $\Sigma_{*}$ for high- and low-mass galaxies. 
It is shown that high-mass group on average has lower local sSFR than 
low-mass group for a spatial region with the same $\Sigma_*$. 

From the integrated SFMS, with a slope value of 1.0, it is known that galaxies with 
higher masses have smaller total sSFR. Here it is shown that higher mass galaxies have smaller 
sSFR in all radii, and it is suggested that they have systematically different spatially-resolved 
SFMS from less-massive galaxies. \citet{perez2013} also show the evidence of systematically smaller 
sSFR of massive galaxies by constructing their spatially resolved stellar mass assembly rates which are 
derived from an analysis of their spatially resolved spectra.
%----------------------------------------------------
% Figure : sSFR and ratio of sSFR
%----------------------------------------------------
%-----------------
% Figure :             

\section{Summary} \label{sec:summary}

We examine the relation between local (1-2 kpc-scale) surface density of 
stellar mass ($\Sigma_*$) and star formation rate ($\Sigma_{\rm SFR}$) by 
analysing the $\Sigma_{*}$ and $\Sigma_{\rm SFR}$ of 93 local massive 
($M_* > 10^{10.5}M_{\odot}$) face-on spiral galaxies located at $0.01<z<0.02$. 
The method so-called pixel-to-pixel SED fitting is used to derive the $\Sigma_{*}$ and $\Sigma_{\rm SFR}$. We find a tight nearly linear relation between local $\Sigma_{*}$ and $\Sigma_{\rm SFR}$ which has monotonic increase (with slope nearly unity) in low $\Sigma_{*}$ (corresponds to the disc region) while nearly flat in high $\Sigma_{*}$ (corresponds to the central region). This relation in a local scale is called spatially-resolved star formation main sequence (SFMS). The spatially-resolved SFMS relation seems to break in quenched galaxies.      
 
Analysis on the systematic offset from a constant sSFR in various scale 
suggested that the scatter in spatially-resolved SFMS is in part due 
to the local variation of sSFR, where spiral arm region has enhanced sSFR 
compared to the underlying disc region, and systematic variation of 
spatially-resolved SFMS in larger scale due to global properties, such 
as $M_*$ and the existence of a bar structure. $M_{*}$ has a role in suppressing 
the local sSFR throughout the galaxy's region in such a way that more 
massive galaxies tend to have local sSFR lower in all radius than 
those of less massive galaxies, hence giving a contribution to the 
scatter in the spatially-resolved SFMS. Barred galaxies in our 
sample have systematically lower sSFR in the core region than non-barred 
galaxies, but their sSFR are similar in the outside regions.

\section*{Acknowledgements}

This work is based on observations made with the NASA \textit{Galaxy Evolution Explorer}. GALEX is operated for NASA by the California Institute of Technology under NASA contract NAS5-98034. This work has made use of SDSS data. Funding for SDSS-III has been provided by the Alfred P. Sloan Foundation, the Participating Institutions, the National Science Foundation, and the U.S. Department of Energy Office of Science. The SDSS-III web site is http://www.sdss3.org/. SDSS-III is managed by the Astrophysical Research Consortium for the Participating Institutions of the SDSS-III Collaboration including the University of Arizona, the Brazilian Participation Group, Brookhaven National Laboratory, Carnegie Mellon University, University of Florida, the French Participation Group, the German Participation Group, Harvard University, the Instituto de Astrofisica de Canarias, the Michigan State/Notre Dame/JINA Participation Group, Johns Hopkins University, Lawrence Berkeley National Laboratory, Max Planck Institute for Astrophysics, Max Planck Institute for Extraterrestrial Physics, New Mexico State University, New York University, Ohio State University, Pennsylvania State University, University of Portsmouth, Princeton University, the Spanish Participation Group, University of Tokyo, University of Utah, Vanderbilt University, University of Virginia, University of Washington, and Yale University.

%%%%%%%%%%%%%%%%%%%%%%%%%%%%%%%%%%%%%%%%%%%%%%%%%%



\bibliographystyle{mnras}
\bibliography{mybib} % if your bibtex file is called example.bib


%%%%%%%%%%%%%%%%%%%%%%%%%%%%%%%%%%%%%%%%%%%%%%%%%%

%%%%%%%%%%%%%%%%% APPENDICES %%%%%%%%%%%%%%%%%%%%%

\appendix

\section{Robustness of our Method} \label{sec:robust}

\begin{figure*}
\centering
\includegraphics[width=1\textwidth]{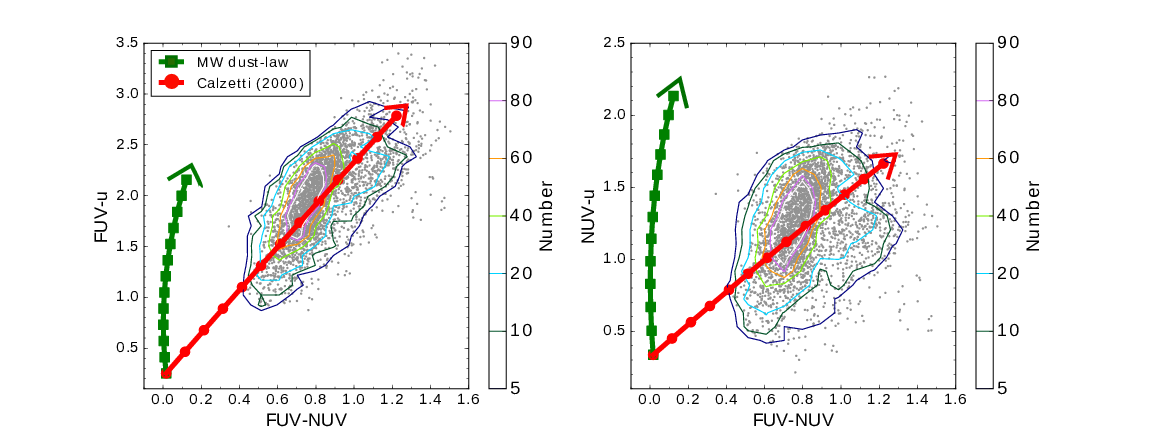}
\caption{Dust extinction tracks of \citet{calzetti2000} (red circles) and Milky Way (green squares) dust extinction laws for a model SED with $Z=0.02$, $\tau=3.7$Gyr, and age=$2.5$Gyr in the colour-colour diagrams (left panel: \textit{FUV}$-$\textit{NUV} vs. \textit{NUV}$-u$, right panel:\textit{FUV}$-$\textit{NUV} vs. \textit{NUV}$-u$) along with the distribution of photometries of the M51's bins. The circles and squares mark increasing colour excess $E(B-V)$ from 0 to 0.6 with a step of 0.05. } 
\label{fig:M51_dustcurve_test}
\end{figure*}

In order to evaluate the reliability of the pixel-to-pixel SED
fitting method in estimating a local SFR, we perform the 
pixel-to-pixel SED fitting in the central region of M51, which have 
resolved MIR dataset from the SINGS survey \citep{kennicutt2003}. 
Figure~\ref{fig:M51_dustcurve_test} shows 
the dust reddening tracks of the \citet{calzetti2000} and Milky-Way dust extinction 
laws in the colour-colour diagram along with the distribution of M51's 
photometric data in the bins. Milky-Way dust law have curvy dust 
extinction tracks in the \textit{FUV}$-$\textit{NUV} vs. \textit{FUV}$-u$ and \textit{FUV}$-$\textit{NUV} vs. \textit{NUV}$-u$ colour-colour 
diagrams which deviate from overall distribution of photometric data,
similar to the results shown in Figure~\ref{fig:Gals_dustcurve_test}. 

\begin{figure*}
\centering
\includegraphics[width=0.9\textwidth]{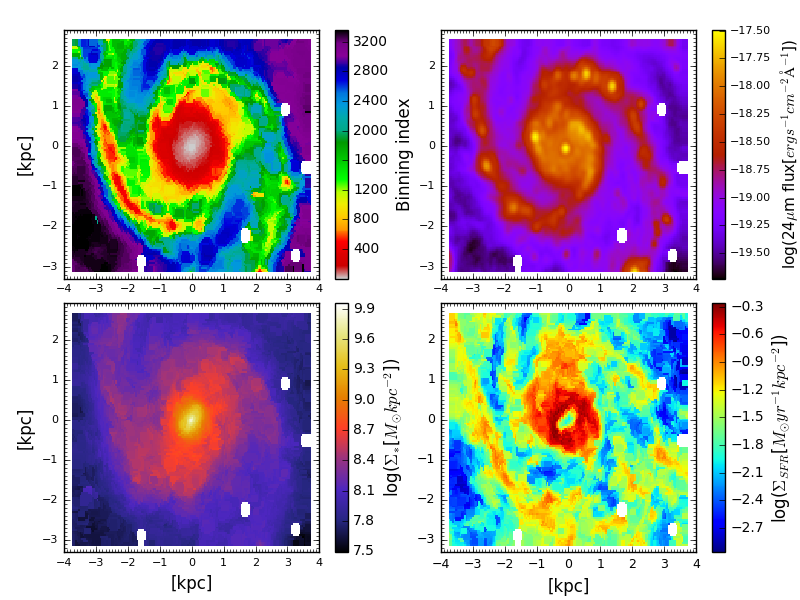}
\caption{Pixel-to-pixel SED fitting result for M51 galaxy. Clockwise from the top-left panel : pixel binning result, spatially-resolved $24\mu m$ flux map, spatially-resolved $\Sigma_{\rm SFR}$ map, and spatially-resolved $\Sigma_{*}$ map.
\label{fig:M51fittingresult}}
\end{figure*}

\begin{figure*}
\centering
\includegraphics[width=1\textwidth]{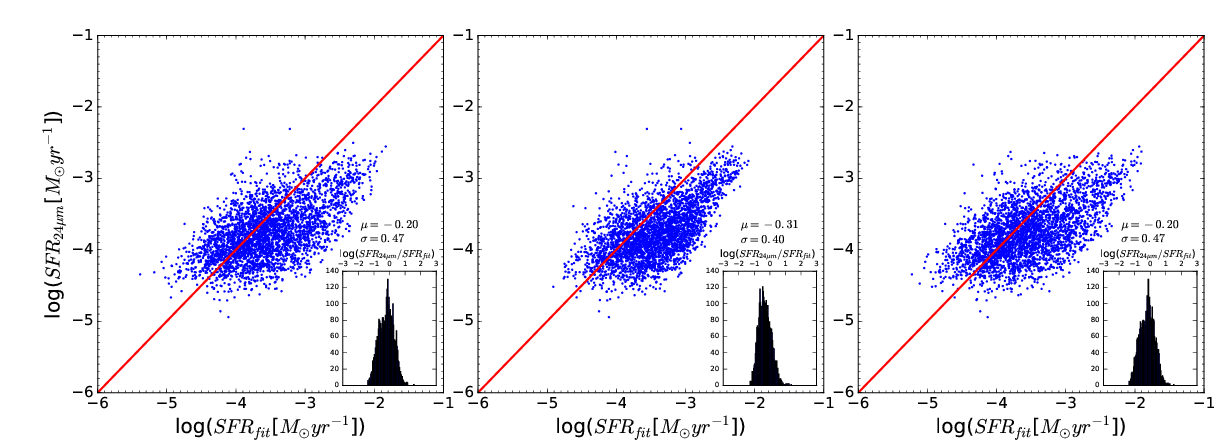}
\caption{Comparison between three fitting method in term of their consistency with SFR from $24\mu m$ flux : minimum $\chi^{2}$ (left panel) and Bayesian statistics-based fitting with two different PDFs, P$(\theta|X)$ following Student's t distribution with $\nu = 3$ (middle panel), and P$(\theta|X)\propto\exp(-\chi^{2}/2)$ (right panel).}
\label{fig:M51SFRfitvs24m_fitmethod}
\end{figure*}

The result of the pixel-to-pixel SED fitting 
is shown in Figure~\ref{fig:M51fittingresult}. We check the reliability of our pixel-to-pixel SED fitting method by comparing the SFR estimates by the maximum likelihood fitting with those by the Bayesian method with two different PDFs. 
Figure~\ref{fig:M51SFRfitvs24m_fitmethod} shows comparison between 
M51's spatially-resolved SFR resulted from the SED fitting. 
The left, middle, and right panels show the results of three different
SED fitting method with the maximum likelihood, Bayesian with $P(\theta|X)$ 
with Student's t distribution with $\nu = 3$, and Bayesian with
$P(\theta|X)\propto \exp(-\chi^{2}/2)$. They are compared with SFR 
estimated from $24\mu$m flux with the SFR-$24\mu$m flux relation in \citet{rieke2009}. We find that the Bayesian method with Student's t distribution with $\nu = 3$ give the best estimate of SFR with smaller scatter from the proportionality. The distribution of the $\log (\text{SFR}_{\rm{fit}}/\text{SFR}_{24\mu m})$ has scatter ($\sigma$) of 0.40 dex and mean value ($\mu$) of $-0.31$ dex.
%----------------------------------------------------
%---------------------------------------------------
%--------------------------------------------------
%---------------------------------------------------
%---------------------------------------------------
% Figure B7 : Fitting method and model weighting
%-------------------------------------------------- 

Our new method of the Bayesian SED fitting with Student's t distribution of $\nu=3$ is also verified by fitting with mock SEDs. We construct mock SEDs generating from \citet{bruzual2003} with randomly selected parameters ($Z$, $\tau$, $E(B-V)$, and age). Some normalizations (randomly selected in the range of $10^{7}\text{to}10^{13}$) are applied to each SED then assigned an S/N ratio. Adding random noise following gaussian distribution with standard deviation correspond to the flux error from the S/N ratio, also applied the Bayesian SED fitting to simulate the observational effect. We fit mock SEDs with Bayesian method with 11 different PDFs: flat (no-weighting), Student's t distribution with $\nu=0.001, 1, 3, 5, 6, 8, 10, 15, 20$, and Gaussian. Each fitting is done with two set of models, one with $\tau$ ranging from $0.1$ to $10$ Gyr and the other with a range of $0.1$ to $2$ Gyr, to test the stability of the method against the choice of model parameter ranges. 

\begin{figure*}
\centering
\includegraphics[width=1\textwidth]{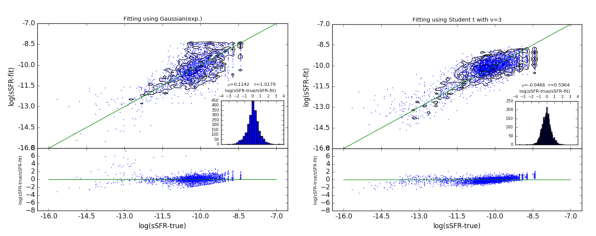}
\caption{Fitting test with mock SEDs using the Bayesian fitting with P$(\theta|X)\propto\exp(-\chi^{2}/2)$ (or gaussian) (left panel) and P$(\theta|X)$ following Student's t distribution with $\nu = 3$ (right panel).}
\label{fig:test_fitting_mockSEDs}
\end{figure*}

\begin{figure*}
\centering
\includegraphics[width=1\textwidth]{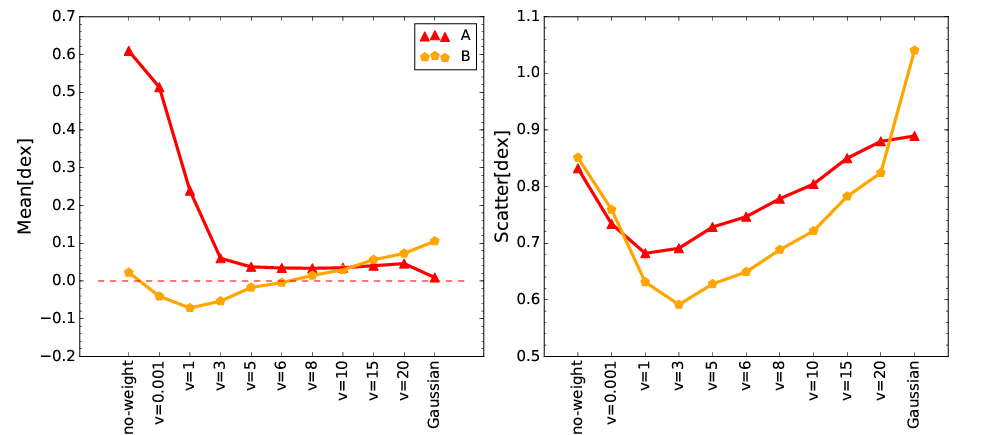}
\caption{Means (left panel) and scatters (right panel) of $\log (\rm SFR-true/\rm SFR-fit)$ distributions from several different fitting methods and mock SEDs. Different symbols and colours mean : mock SEDs with random noise fitted to model SEDs with $\tau[0.1:2]$ (A) and $\tau[0.1:10]$ (B). x-axis represent Bayesian fitting method.}
\label{fig:test_fitting_mockSEDs_compile}
\end{figure*}

Figure~\ref{fig:test_fitting_mockSEDs} shows example of two fitting results, mock SEDs with $\tau$ from $0.1$ to $10$ Gyr using Bayesian method with Gaussian PDFs (left panel) and Student's t distribution of $\nu=3$ (right panel). Histogram in the inset shows distributions of $\log (\rm SFR-true/\rm SFR-fit)$. The distributions indicate that Bayesian method with PDF of Student's t distribution of $\nu=3$ is better than those with Gaussian PDFs as it has lower systematic offset and scatter. Figure~\ref{fig:test_fitting_mockSEDs_compile} shows means (left panel) and scatters (right panel) of $\log (\rm SFR-true/\rm SFR-fit)$ from various PDFs. The x-axis represents a various Bayesian fitting method with $\nu$ indicate the $\nu$ value of Student's t distribution. Different symbols and colours mean : mock SEDs with random noise fitted to model SEDs with $\tau[0.1:2]$ (red triangles) and $\tau[0.1:10]$ (yellow pentagons). The Bayesian method with PDF of Student's t distribution with $\nu=3$ gives the best result compared to the other PDFs in the fitting tests.                

%---------------------------------------------------
% figure : test fitting with mock SEDs :
%---------------------------------------------------
%--------------------------------------------------
% figure : test fitting with mock SEDs compilation :
%--------------------------------------------------
%%%%%%%%%%%%%%%%%%%%%%%%%%%%%%%%%%%%%%%%%%%%%%%%%%

% Don't change these lines
\bsp	% typesetting comment
\label{lastpage}
\end{document}